\documentclass[final,5p,times,twocolumn]{elsarticle}

\usepackage{amssymb}
\usepackage{color}
\usepackage{amsmath}
\usepackage{multirow}
\usepackage{booktabs}
\usepackage{bm}
\usepackage{braket}
\usepackage{amsmath}
\usepackage{graphicx}
\usepackage {ulem}
\biboptions{sort&compress}

\usepackage[pdfstartview=FitH, colorlinks,
 linkcolor={red!60!black!},
 anchorcolor={green!80!black!},
 urlcolor={blue!80!black!},
 citecolor={blue!50!black!}]{hyperref}
\usepackage[table]{xcolor}

\journal{Physics Letters B}

\begin{document}


\begin{frontmatter}




\title{Spectroscopic factors of resonance states with the Gamow shell model}


\author[ad1,ad2]{M.R. Xie}
\author[ad1,ad2]{J.G. Li \corref{correspondence}}
\author[ad1,ad2]{N. Michel}
\author[ad1,ad2]{H.H. Li}
\author[ad1,ad2]{W. Zuo}

\address[ad1]{CAS Key Laboratory of High Precision Nuclear Spectroscopy, Institute of Modern Physics,
Chinese Academy of Sciences, Lanzhou 730000, China}
\address[ad2]{School of Nuclear Science and Technology, University of Chinese Academy of Sciences, Beijing 100049, China}

\cortext[correspondence]{Corresponding author:  jianguo\_li@impcas.ac.cn (J.G. Li)}

\begin{abstract}
We provide an investigation of the spectroscopic factor of resonance states in $A =5-8$ nuclei, utilizing the Gamow shell model (GSM).
Within the GSM, the configuration mixing is taken into account exactly with the shell model framework, and the continuum coupling is addressed via the complex-energy Berggren ensemble, which treats bound, resonance, and non-resonant continuum single-particle states on an equal footing.  As a result, both the configuration mixing and continuum coupling are meticulously considered in the GSM.
We first calculate the low-lying states of helium isotopes and isotones with the GSM, and the results are compared with that of \textit{ab initio} no-core shell model (NCSM) calculations. The results indicate that GSM can reproduce the low-lying resonance states more accurately than the no-core shell model.
Following this, we delve into the spectroscopic factors of the resonance states as computed through both GSM and NCSM, concurrently conducting systematic calculations of overlap functions pertinent to these resonance states.
Finally, the calculated overlap function and spectroscopic factor of $^6$He$(0_1^+)$ $\otimes \nu p_{3/2} \to $ $^7$He$(3/2_1^-)$ with GSM are compared with the results from \textit{ab initio} NCSM, variational Monte Carlo, and Green's function Monte Carlo calculations, as well as avaliable experimental data. 
The results assert that wave function asymptotes can only be reproduced in GSM, where resonance and continuum coupling are precisely addressed.

\end{abstract}


\end{frontmatter}

\section{Introduction.}

Large scientific facilities have been upgraded or are being constructed, such as 
the facility for rare isotope beams (FRIB), RI Beam Factory (RIBF), and High-Intensity Heavy-ion Accelerator Facility (HIAF), which have advanced the study of exotic nuclei away from the valley of stability and propelled it to the forefront of nuclear physics research.
The peculiar phenomena that distinguish unstable nuclei from those close to the valley of stability include shell evolution \cite{SMIRNOVA2010109,PhysRevLett.109.032502}, halo and skin nuclei \cite{RevModPhys.76.215,PhysRevLett.116.212501}, exotic particle emission \cite{RevModPhys.84.567,DETRAZ1980307,KOURA2015228,PhysRevLett.43.1652}, etc. Such phenomena occur as a consequence of valence nucleons primarily occupying shells near the particle-emission threshold. The study of the structure of exotic nuclei has generated great interest both theoretically and experimentally.
Many works have already been done on these subjects, particularly about halo nuclei \cite{PhysRevC.64.061301,BAYE2011464,PhysRevLett.124.212503,PhysRevLett.112.242501}. However, some questions remain unanswered  regarding the coupling of resonance and scattering states \cite{PhysRevLett.89.042501}, while relevant experimental data are lacking or even unavailable. 
Resonance nuclei located far from the valley of stability and close to or beyond the drip line form a basic laboratory to study the coupling of discrete states with scattering continuum states \cite{PhysRevC.105.L051301}. 
Traditional shell models have difficulties in describing the properties of these resonance nuclei. It necessitates the introduction of new and powerful models that account for the complicated interplay among the different degrees of freedom in those weakly-bound and resonance states \cite{PhysRevC.72.054322,PhysRevC.67.054311}.



A typicaly feature of many-body resonance states is the strong coupling to continuum.
The coupling between the resonance states and the continuum poses the greatest challenges in the theoretical description of unbound nuclei. 
Various studies have demonstrated that continuum coupling has a significant impact on the properties of the nucleus in the drip line regions, such as energy spectra \cite{PhysRevC.100.064303,ZHANG2022136958,PhysRevC.104.024319,PhysRevC.103.034305,PhysRevLett.127.262502}, asymptotic normalization coefficient (ANC) \cite{PhysRevC.85.064320}, spectroscopic factor (SF) \cite{PhysRevC.75.031301,PhysRevC.82.044315,XIE2023137800,PhysRevC.104.L061301}, etc. 


In nuclear structure studies, the shell model embedded in the continuum (SMEC) \cite{PhysRevC.67.054311, OKOLOWICZ2003271}, which treats the bound states and the scattering states on an equal footing, has been developed these recent years. 
While it is in principle possible to include two particles in the continuum in SMEC, as in the context of two-proton radioactivity \cite{Blank_2008, PhysRevLett.95.042503}, it can only be performed with approximations, such as cluster and sequential emissions.
In contrast, the Gamow shell model (GSM) used in the present work effectively avoids this defect and allows multiple particles to be present in the non-resonant continuum.
GSM allows to perform many-body computations using the Gamow-Berggren basis in rigged Hilbert spaces \cite{BERGGREN1968265,0954-3899-36-1-013101,physics3040062, Michel_Springer}.
The Berggren basis extends the Schr\"{o}dinger equation to the complex momentum space, whereby the scattering contours associated with resonance states are discretized with the Gauss-Legendre rule. Physical quantities are complex so the width of resonance states can be obtained from the imaginary part of energy.
A variety of phenomena induced by the proximity of the nucleon-emission threshold shows that the asymptotic part of many-body wave functions must be precisely reproduced, which is the case in GSM.
GSM has thus been used to investigate unbound states in drip line nuclei, such as the conditions of the existence of a $4n$ resonance system \cite{PhysRevC.100.054313}, the prediction of the ground-state energy of $^{28}$O \cite{PhysRevC.103.034305}, and the calculations of energies and widths associated to the one-proton and two-proton decays of $^{18}$Mg \cite{PhysRevC.103.044319, PhysRevLett.127.262502} and to the possibility of a narrow resonance ground state in $^7$H \cite{PhysRevC.104.L061306}.



SF and its associated overlap function are useful tools for the study of the structure of many-body wave functions of unbound nuclei.
SF represents the amount of occupancy of a specific single-particle (s.p.) orbit and associated shell model configuration in the many-body nuclear wave function.
It is also a physical quantity that connects the nuclear structure to the nuclear reaction~ \cite{PhysRevLett.95.222501,doi:10.1146/annurev.nucl.53.041002.110406, PhysRevC.105.024613}. 
However, at present, theoretical SFs differ significantly from experimental data~ \cite{PhysRevC.90.057602,PhysRevC.103.054610,PhysRevC.105.024613,PhysRevC.104.L061301}. 
Studies have also demonstrated that continuum coupling provides large contributions to SFs in resonance and weakly-bound nuclei \cite{PhysRevC.104.L061301, XIE2023137800}. As both continuum effects and inter-nucleon correlations are taken into account simultaneously in the GSM \cite{PhysRevC.70.064313}, the asymptotes of many-body systems are precisely evaluated. 
\textit{Ab initio} no-core shell model (NCSM) is also considered for comparison, in which all nucleons are active and interact via a realistic interaction \cite{BARRETT2013131}. 
And the basis of the harmonic oscillator (HO) state is used, in which only bound valence shells are mainly occupied in the s.p.~basis. 

After a brief introduction to the theoretical frameworks of GSM, we compare the results of calculated energy spectra and resonance widths of neutron-rich helium isotopes and their proton-rich isotones with experimental data.
SFs and overlap functions of resonance states will be calculated using both GSM and NCSM. Results allow us to investigate the effects of continuum coupling in resonance states. A summary is provided afterwards.

\section{Method.}
The Gamow state, also known as the complex-energy resonance state, was firstly proposed by George Gamow in 1932 to describe $\alpha$ decay \cite{PhysRev.56.750}.
By describing a nuclear state in the complex-energy plane, the theory of Gamow resonance state provides a natural definition of resonance state \cite{PhysRev.56.750}.
Resonance states lie above decay thresholds so that they can decay via the emission of a nucleon, or of clusters such as deuteron, $^3$He/triton, $\alpha$ particles, etc.
For a given GSM eigenstate, the eigenenergy of the many-body state can be written as~\cite{PhysRevLett.89.042502},
\begin{equation}
    \tilde{E}=E-i\Gamma/2, 
\end{equation}
where the real part $E$ corresponds to the position of the resonance, while the imaginary part $\Gamma$ is the resonance width, which is related to the lifetime of the state.



The Berggren completeness relation allows to include resonance states in a complete set of one-body states. It was first formulated by Tore Berggren and its expression is as follows \cite{BERGGREN1968265} :
\begin{equation}
    \sum_n u_n(r)u_n(r')+\int_{L^+} u(k,r) u(k,r') dk = \delta(r-r'), 
\label{Berggren}
\end{equation}
where $u_n(r)$ is a bound or resonance state and $u(k,r)$ is a scattering state belonging to the $L_+$ contour in the complex-momentum plane.
Within the Berggren representation, continuous states are located on the contour $L_{+}$ of the fourth quadrant of the complex plane,
and the resonance states lying between the contour $L_{+}$ and the real-energy axis are included in the discrete sum of Eq.~(\ref{Berggren}) along with bound states.
Continuous states on the contour $L_+$ are discretized using the Gauss-Legendre method.

Due to the presence of complex-energy states in the Berggren basis of Eq.~(\ref{Berggren}), the Hamiltonian of GSM is represented therein by a complex symmetric matrix,
which is diagonalized using the complex extension of the Jacobi-Davidson method \cite{Michel_Springer}.
Resonance states are identified in the complex-energy eigenspectrum of unbound states formed by both resonant and scattering eigenstates by using the overlap method \cite{0954-3899-36-1-013101,Michel_Springer}. 

%

The overlap function $I_{\ell j}(r)$ is defined as:
\begin{eqnarray}
\label{overlap}
I_{\ell j}(r) & = & \langle \Psi_A^{J_A}| \left [|\Psi_{A-1}^{J_{A-1}}\rangle \otimes |r \ell j\rangle^{J_A} \right]\rangle, \nonumber \\
          & = & \frac{1}{\sqrt{2J_A + 1}} \sum_i \langle \Psi_A^{J_A}||a_{n_i \ell j}^{\dagger}|| \Psi_{A-1}^{J_{A-1}}\rangle  u_{n_i}(r),
\end{eqnarray}
where $|\Psi^{J_A}_A\rangle$ and $|\Psi^{J_{A-1}}_{A-1}\rangle$ are the wave functions of the $A$ and $A-1$ nuclear systems, respectively;
$\ell$ and $j$ are respectively the orbital and total angular momenta of the considered partial wave in $I_{\ell j}(r)$;
$a_{n_i \ell j}^{\dagger}$ is the creation operator associated with the $|n_i \ell j \rangle$ state and is either the $n_i$-th state of the HO basis (NCSM) or of the discretized Berggren basis (GSM);
$u_{n_i}(r)$ represents the radial wave function of the $|n_i \ell j \rangle$ state.

SF is defined as being the norm of the radial overlap function $I_{\ell j}(r)$ of Eq.~(\ref{overlap}):

\begin{equation}
\label{specfac}
C^2 S = \int I_{\ell j}(r)^2 dr,
\end{equation} 
where $C^2 S$ is a standard notation for SF.
In GSM, the imaginary parts of SFs can be interpreted as the uncertainty on real parts  \cite{0954-3899-36-1-013101,Michel_Springer}. 
From Eqs.~(\ref{overlap},\ref{specfac}), the complex effects of inter-nucleon correlations and continuum coupling on the asymptotic behavior of resonance states can be directly pointed out \cite{PhysRevC.75.031301,PhysRevC.104.L061301}.

\section{Results.}

\begin{figure}
    \centering
    \includegraphics[width=0.4\paperwidth]{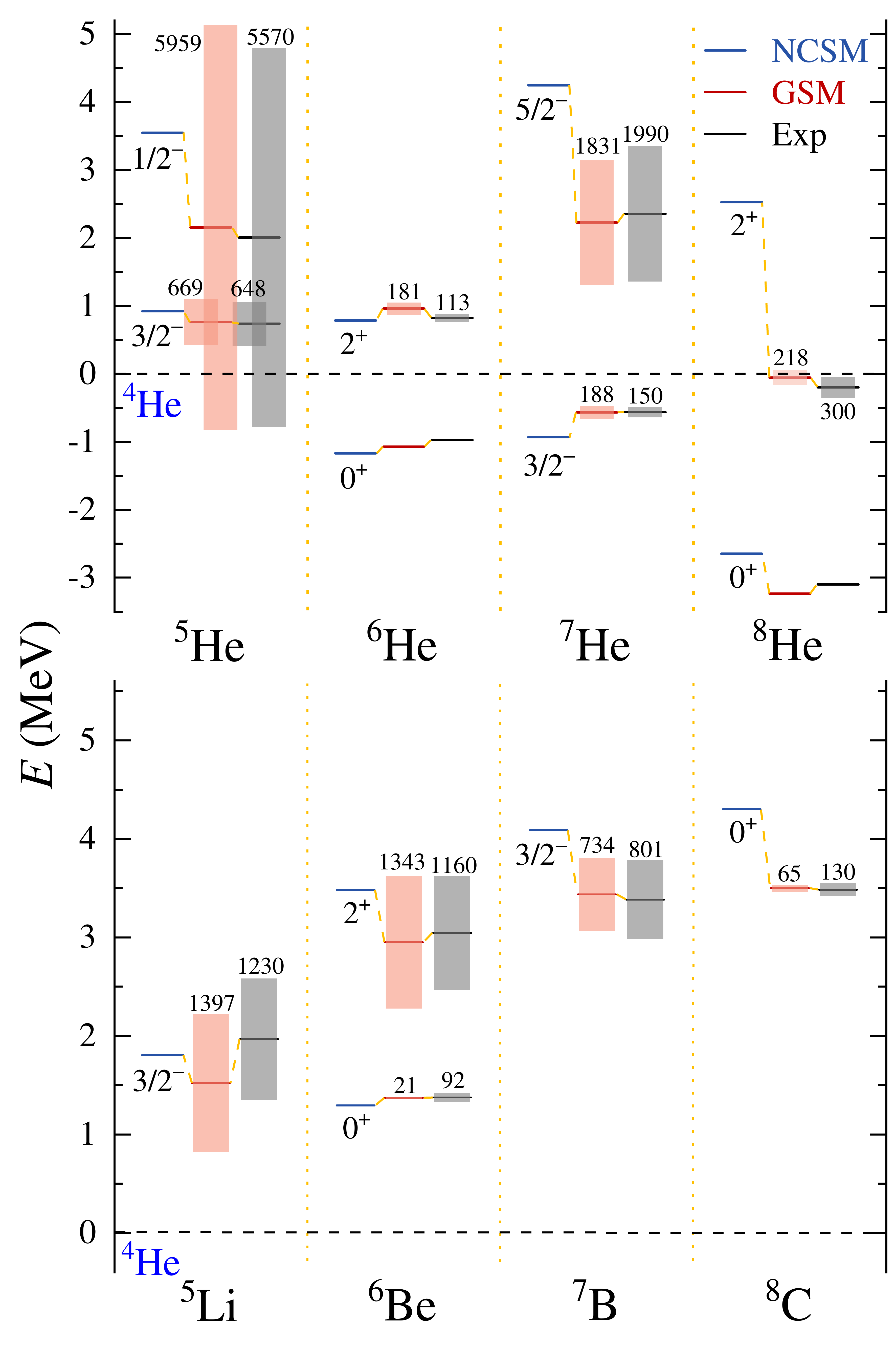}
    \caption{Low-lying spectra (in MeV) and widths (in keV) of the isotopes of $^{5-8}$He (upper), and the isotones of $^{5}$Li, $^{6}$Be, $^{7}$B, and $^{8}$C (lower). The calculations illustrated by blue and red lines are provided by NCSM and GSM, respectively, and the experimental data are taken from Ref. \cite{ensdf,TILLEY20023,PhysRevC.84.014320,PhysRevC.82.041304,PhysRevC.84.014320}. Ground-state energies are given with respect to that of $^4$He. Shaded areas represent the resonance width in the GSM calculations and experimental data.}
    \label{energy}
\end{figure}

We use GSM in the core plus valence particle picture.
The GSM Hamiltonian consist of a one-body finite depth Woods-Saxon (WS) potential mimicking the inert $^4$He core and of a residual two-body interaction among valence nucleons.
In the present work, we use the Gaussian-based Furutani, Horiuchi, and Tamagaki (FHT) residual two-body interaction \cite{10.1143/PTP.62.981}.
The Hamiltonian parameters and model space of Ref. \cite{PhysRevC.96.054316} are adopted.
The FHT interaction has been successfully used in the GSM to describe the properties of $A=5- 10$ light neutron-rich nuclei \cite{PhysRevC.96.054316,PhysRevC.104.L061306}, the mirror states in $^{16}$F/$^{16}$N \cite{PhysRevC.106.L011301}, and radiative capture reactions of $A = 6 - 8$ nuclei  \cite{PhysRevC.91.034609}.
We only consider weakly bound and resonance $p$-shell nuclei.
All partial waves up to $l$ = 3 are taken into account.
Among them, the $d_{3/2}$, $f_{7/2}$, and $f_{5/2}$ partial waves bear large centrifugal barriers, and their influence on the asymptotic behavior of the wave function is negligible, so they are expanded on HO basis. While the partial waves of $s_{1/2},p_{3/2},p_{1/2}$, and $d_{5/2}$ use Berggren basis to represent.
The $p_{3/2}$ and $p_{1/2}$ partial waves are discretized by 60 points along the contour $L_{+}$, while the $s_{1/2}$ and $d_{5/2}$ partial wave are discretized by 30 points.
The maximal momentum value for the contours is $k_{\max } = 4.0$ $\mathrm{fm}^{-1}$.
In performed GSM calculations, three nucleons at most can occupy states in the non-resonant continuum.
We also performed NCSM calculations for comparison. Within the NCSM calculations, we use the Daejeon16 two-body interaction, which is based on an SRG-transformed chiral N$^3$LO interaction \cite{SHIROKOV201687}. It provides a good description of various observables in light nuclei without three-body forces and generates rapid energy convergence in \textit{ab initio} calculations \cite{SHIROKOV201687}. The HO basis with a $\hbar \omega = 15$ MeV frequency is utilized. 
Due to limited computing power, maximal truncations are $N_{\rm max}$ = 12 for $A$ = 5 and 6 nuclei, and $N_{\rm max}$ = 10 for $A$ = 7 and 8 nuclei.
The results of extrapolating the energy spectra to infinite-size model spaces \cite{PhysRevC.99.054308, Shin_2017}.


The energies of the low-lying states of the $^{5-8}$He isotopes and their isotones are calculated in NCSM and GSM.
Results are displayed in Fig.~\ref{energy}, along with experimental data \cite{ensdf,TILLEY20023,PhysRevC.84.014320,PhysRevC.82.041304,PhysRevC.84.014320}. 
As one considers mirror nuclei, comparing their energies and decay widths is interesting, which would be identical if the isospin symmetry is strictly conserved.
Indeed, significant differences appear due to the presence of the Coulomb Hamiltonian among valence protons in helium isotones.
It is noted that the ground-state energies of both helium isotopes and isotones calculated using NCSM and GSM are close to experimental data. However, the excitation energy of the excited state in NCSM is usually higher than that of GSM and experimental data.
As the Coulomb barrier is low in light nuclei, unbound states of sizable width can form at the proton drip line.
GSM accurately reproduces the proton-emission widths of the resonance states of helium isotopes and isotones. The difference between calculated and experimental widths is indeed smaller than 200 keV.
According to our GSM calculations, $^5$He($1/2^-$) is a broad ($\Gamma$ = 5959 keV) resonance, as well as the $^7$He($5/2^-$) excited state ($\Gamma$ = 1831 keV), values which are both consistent with experimental data, 5570 and 1990 keV for $^5$He($1/2^-$) and $^7$He($5/2^-$) \cite{TILLEY20023}, respectively.

The NCSM is diagonalized using a basis of HO states, implying that the nucleus is viewed as a CQS. While such an assumption is justified for well-bound nuclei, it can no longer be applied to unbound states.
Weakly bound and unbound states bear a low-lying  emission threshold and a large surface density space dispersion due to strong coupling to scattering states. 
It can then be expected that GSM will be better at describing resonance state features than NCSM.

\begin{figure*}[!htp]
    \centering
    \includegraphics[width=0.88\paperwidth]{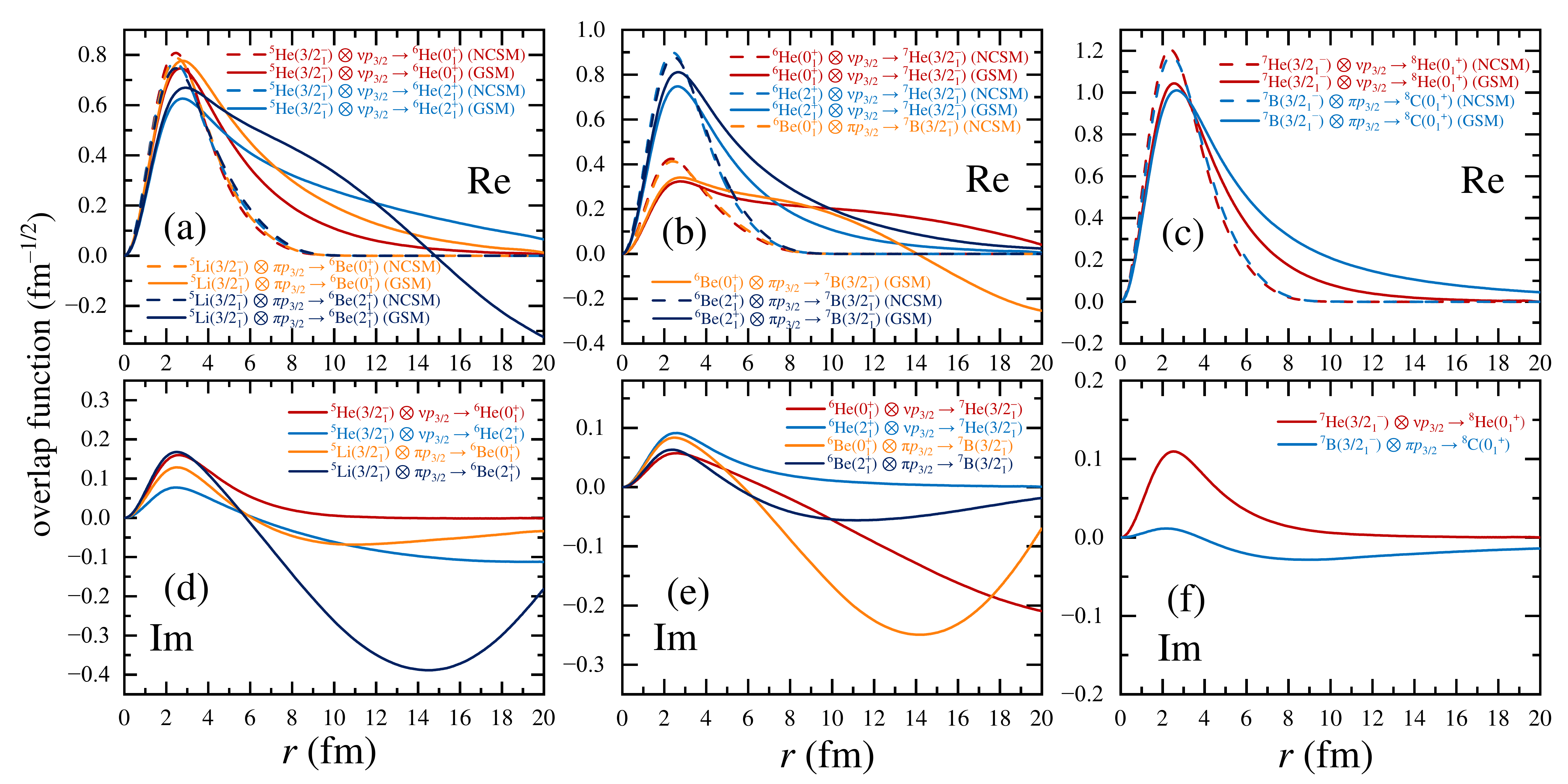}
    \caption{Calculated the overlap functions of $p_{3/2}$ partial wave in $\ket{A=5} \otimes p_{3/2} \rightarrow \ket{A=6}$, $\ket{A=6} \otimes p_{3/2} \rightarrow \ket{A=7}$, and $\ket{A=7} \otimes p_{3/2} \rightarrow \ket{A=8}$  systems with NCSM (dashed lines) and GSM (solid lines). The real part of NCSM and GSM are shown in the upper panel, and the imaginary part of GSM is shown in the lower panel. 
    }
    \label{overlap function}
\end{figure*}

To explore the properties of resonance states more comprehensively, particularly the asymptotic nature of the wave function, it is crucial to conduct further investigations. Resonance states are characterized by their temporary nature and their wave functions exhibit distinct behavior in the asymptotic region, which is crucial for understanding their decay properties and scattering phenomena. Next, we calculated the overlap functions and SFs ($C^2S$ in this paper) of the associated resonance states using GSM.
$C^2S$ values and overlap functions in He, Li, Be, B, and C isotopes bearing $A = 6 - 8$ nucleons are calculated, and the results are presented in Table \ref{c2s_1} and Fig. \ref{overlap function}.
The NCSM calculations are also performed for comparison. 
The convergence of $C^2S$ is slower as model spaces increase in dimension compared to the energy in NCSM calculations.
Consequently, it is necessary to consider the $C^2S$ and overlap functions provided by the largest NCSM model space for comparison with GSM calculations, as presented in Table \ref{c2s_1} and Fig. \ref{overlap function}.

\begin{center}
\begin{table}[]
\caption{$C^2S$ values issued from NCSM and GSM calculations for $p_{3/2}$ and $p_{1/2}$ partial waves.  One uses a NCSM model space of $(N_{\rm max}^{A-1},N_{\rm max}^A)=(12,12)$ for $A=$ 5 and 6 nuclei, and  $(N_{\rm max}^{A-1},N_{\rm max}^A)=(10,10)$ for $A=$ 7 and 8 nuclei. Parent and daughter nuclei are denoted by $A-1$ and $A$, respectively.}
\label{c2s_1}
\resizebox{0.49\textwidth}{!}{
\begin{tabular}{cccccccc}
\hline \hline
\multicolumn{5}{c}{} & \multicolumn{2}{c}{$C^2S$}    \\ 
  $A-1$ & $(J^{\pi})_{A-1}$ & $A$ & $(J^\pi)_A$ & Wave & GSM & NCSM \\ \hline \hline
    $^5$He&$3/2^-_1$ &$^6$He& $0^+_1$ & $p_{3/2}$&$1.850-i0.747$&1.572 \\
      &$3/2^-_1$  &  &$2^+_1$& $p_{3/2}$&$1.855+i0.096$&1.490\\
      &$1/2^-_1$&  &$2^+_1$ & $p_{3/2}$&$0.138-i0.015$&0.094 \\ 
      &$1/2^-_1$&  &$0^+_1$ & $p_{1/2}$&$0.246-i0.150$&0.104 \\ \hline
      
$^6$He&$0^+_1$ &$^7$He&$3/2^-_1$& $p_{3/2}$&$0.487+i0.101$&0.455 \\
 &$0^+_1$ & &$1/2^-_1$& $p_{1/2}$&$0.585-i0.023$&0.712 \\
 &$2^+_1$ & &$3/2^-_1$& $p_{3/2}$&$1.865-i0.446$&1.931 \\

 & $2^+_1$ & &$5/2^-_1$& $p_{3/2}$&$0.140-i0.070$&0.208 \\ 
 & $2^+_1$ & &$5/2^-_1$& $p_{1/2}$&$0.821-i0.005$&0.681 \\\hline
 
$^7$He&$3/2^-_1$ &$^8$He& $0^+_1$ & $p_{3/2}$&$3.222-i0.664$&3.384 \\
&$3/2^-_1$ &      &$2^+_1$ & $p_{1/2}$&$0.996-i0.127$&0.788 \\
      &$5/2^-_1$ &      &$2^+_1$ & $p_{3/2}$&$2.386+i0.788$&1.724 \\ \hline
      
$^5$Li&$3/2^-_1$ &$^6$Be&$0^+_1$ & $p_{3/2}$&$2.421-i0.353$&1.540 \\
      &  $3/2^-_1$        &      &$2^+_1$ & $p_{3/2}$&$1.822+i0.022$&1.461 \\ \hline
      
$^6$Be&$0^+_1$&$^7$B& $3/2^-_1$ & $p_{3/2}$&$0.436+i0.012$&0.450 \\
      &$2^+_1$&      &    & $p_{3/2}$&$2.573-i0.074$&1.906 \\   \hline
      
$^7$B &$3/2^-_1$ &$^8$C  &$0^+_1$ & $p_{3/2}$&$3.788+i0.112$&3.340\\

\hline \hline
\end{tabular}}
\end{table}
\end{center}

Our investigation reveals that in certain instances, the $C^2S$ derived from both the GSM and NCSM closely. Specifically, in cases such as $^6$He$(0_1^+)$ $\otimes \nu p_{3/2} \to $ $^7$He$(3/2_1^-)$, $^6$He$(2_1^+)$ $\otimes \nu p_{3/2} \to $ $^7$He$(3/2_1^-)$, and $^7$He$(3/2_1^-)$ $\otimes \nu p_{3/2} \to $ $^8$He$(0_1^+)$, both models yield nearly identical results(see Table~\ref{c2s_1} for detail). This uniformity can be attributed to the fact that the correlated nucleon states of $^{6-8}$He are bound or narrow resonance states in these cases. 
However, it is important to note that this alignment is not always observed. In a majority of the cases, a significant discrepancy between the results derived from GSM and NCSM is seen. 
The GSM, in contrast to the NCSM, includes the coupling to the continuum. This refers to the effect of unbound single-particle states, which form a continuum of energies rather than discrete energy spectra. In NCSM, the HO basis is used and does not consider this continuum coupling. It has been shown that continuum coupling is important in describing weakly bound and unbound nuclei, where the coupling to the continuum can significantly affect observables~\cite{PhysRevC.67.054311, PhysRevC.103.034305, PhysRevC.85.064320, XIE2023137800}. 
For instance, when the $A-1$-nucleus is $^5$Li (g.s.) or the $A$-nucleus is $^7$He($5/2^-_1$), which are broad resonances, this discrepancy is particularly apparent. The largest difference between the results of NCSM and GSM calculations is indeed observed for the reaction $^5$Li($3/2^-_1$) $\otimes \pi p{3/2} \to $ $^6$Be($0^+_1$), as well as the reactions $^6$He($2^+_1$) $\otimes \nu p{3/2} \to $ $^7$He($5/2^-_1$) and $^7$He($5/2^-_1$) $\otimes \nu p{1/2} \to $ $^8$He($2^+_1$). 


Overlap functions can provide details on the nuclear structure both inside the nucleus and in the asymptotic zone \cite{LI2022137225, PhysRevC.85.064320}.
They are illustrated in Fig.~\ref{overlap function}, where
both real and imaginary parts are shown in GSM.
We only considered overlap functions involving the $p_{3/2}$ partial wave because the latter is dominant in the studied nuclei.
As the centrifugal barrier is small in the $\ell = 1$ partial wave, one can expect that the overlap function will expand in the asymptotic region.
This is indeed the case for all the overlap functions obtained in GSM, which always lie above NCSM overlap functions in the asymptotic region (see  Fig.~\ref{overlap function}).
This phenomenon is even more pronounced when the state of the $A$ or $A-1$ nucleus lies beyond the particle-emission threshold.
Indeed, due to the unbound character of the involved wave function, the overlap function can present both negative values in its real part and large imaginary parts in modulus,
whereby an oscillation pattern develops (see the overlap functions of the reaction $^5$Li($3/2^-_1$) $\otimes \nu p_{3/2} \to $ $^6$Be($2^+_1$), $^6$Be($0^+_1$) $\otimes \pi p_{3/2} \to $ $^7$B($3/2^-_1$), and $^6$He($0^+_1$) $\otimes \nu p_{3/2} \to $ $^7$He($3/2^-_1$) in Fig.~\ref{overlap function}). 
Conversely, the overlap functions obtained in the NCSM are large inside the nucleus and rapidly decrease to zero outside the nucleus, which is caused by the localized character of the HO basis.

\begin{figure}[!htp]
    \centering
   \includegraphics[width=0.4 \paperwidth]{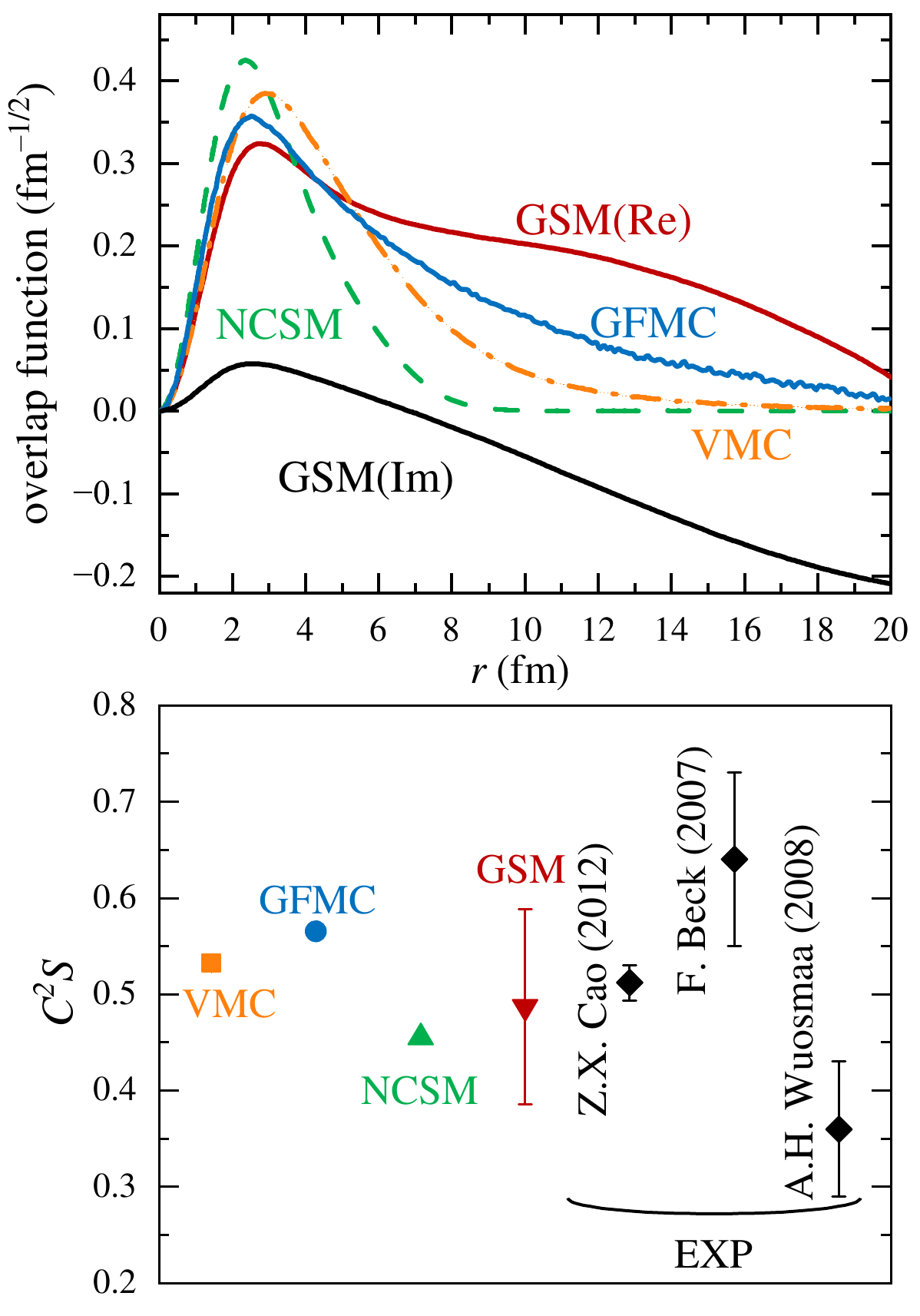}
    \caption{Comparison of the calculated overlap function and $C^2S$ of $^6$He$(0_1^+)$ $\otimes \nu p_{3/2} \to $ $^7$He$(3/2_1^-)$ using GSM and NCSM with VMC and GFMC calculations \cite{PhysRevC.84.024319}, as well as experimental data ~\cite{CAO201246,BECK2009496,PhysRevC.78.041302}. The imaginary part of the GSM is also shown.}
    \label{S}
\end{figure}

Drip line nuclei typically exhibit significant isospin symmetry breaking, such as large mirror energy difference of mirror states \cite{PhysRevLett.125.192503, PhysRevLett.109.102501} and abnormal isobaric multiplet mass equation \cite{PhysRevLett.109.102501}.
The GSM overlap function shows a significant isospin symmetry breaking in investigated mirror partners.
For instance, the asymptotic behavior of the nuclei in the reaction $^5$He($3/2^-_1$) $\otimes \nu p_{3/2} \to $ $^6$He($2^+_1$) and in its mirror case, as shown in Fig.~\ref{overlap function} (a), show noticeable differences.
Moreover, the values of the overlap function associated with the reaction $^5$He($3/2^-_1$) $\otimes \nu p_{3/2} \to $ $^6$He($2^+_1$) are smaller than that of its mirror inside of the nucleus.
This situation reverses outside of the nucleus, whereby the overlap function of the mirror reaction $^5$Li($3/2^-_1$) $\otimes \pi p_{3/2} \to $ $^6$Be($2^+_1$) rapidly decreases therein and can become negative. The imaginary parts of overlap functions are also larger in the mirror reaction of the proton-rich side, due to the larger widths present therein in the $A$ or $A-1$ nucleus.
A similar phenomenon of isospin symmetry breaking can also be seen in the GSM calculations illustrated in Fig.~\ref{overlap function} (b) and (c).
On the contrary, the difference between mirror nuclei is barely discernible in NCSM overlap functions, even though the Coulomb interaction is considered exactly in NCSM (see Fig.~\ref{overlap function}).
Thus, the influence of continuum coupling in the observables involving drip line nuclei is more significant for isospin symmetry breaking. 
In fact, only GSM can depict continuum coupling properly via the explicit use of the Berggren basis.
While the effects of the non-resonant continuum can be mimicked by a few HO basis states in energy spectra, this is not the case for observables depending on the structure of nuclear wave functions in the asymptotic region, such as $C^2S$ and overlap functions.

Finally, we compare the $C^2S$ value and overlap functions of $^6$He$(0_1^+)$ $\otimes \nu p_{3/2} \to $ $^7$He$(3/2_1^-)$ from GSM  with \textit{ab} initio NCSM, variational Monte Carlo (VMC) and Green's function Monte Carlo (GFMC) \cite{PhysRevC.84.024319} calculations, because of the $C^2S$ value has been measured experimentally \cite{CAO201246, BECK2009496, PhysRevC.78.041302}. The results are presented in Fig. \ref{S}.
Due to the wave function of GSM being a complex value, the imaginary part of the overlap function of GSM is presented in Fig. \ref{S}.
Moreover, the calculated $C^2S$ are expressed with an associated error, which originates from the imaginary part of the computed $C^2S$.
Results indicate that the overlap functions calculated with GSM and GFMC are more extended than those from NCSM and VMC and that the GSM overlap function exhibits the resonance character of $^{7}$He($3/2^-$) state. Note that the $^7$He is a resonance, which is well described in GSM calculations, providing insights into the behavior of wave functions as they approach the asymptotic region. However, the $C^2S$ values obtained from VMC and GFMC are slightly larger than those from NCSM and GSM. Nevertheless, Experiment provides an SF of 0.512(18) for $^6$He$(0_1^+)$ $\otimes \nu p_{3/2} \to $ $^7$He$(3/2_1^-)$ \cite{CAO201246}, all calculated $C^2S$ values are in agreement with the experimental data \cite{CAO201246}.

\section{Summary.}

The study of unbound resonance states is of great significance for the comprehension of the properties of atomic nuclei at drip lines. 
We investigated neutron-rich nuclei and their proton-rich mirrors at neutron and proton drip lines using the GSM.
The advantage of GSM to describe resonance states could be precisely assessed by comparing nuclear energies, widths, as well as SFs, and overlap functions with that of \textit{ab initio} no-core shell model calculations.

First, we calculated the energy spectra of the helium isotopes and isotones with GSM, where the decay widths of resonance states are also provided from GSM calculations.
GSM fairly describes the low-lying energy spectra and resonance widths of $^{5-8}$He isotopes and $^5$Li, $^6$Be, $^7$B, and $^8$C isotones.
Compared to NCSM results, we could  see that continuum effects play an essential role in the description of weakly bound and resonance states. 
The asymptotic behavior of resonance states could be precisely investigated in nuclear many-body wave functions from the calculation of SFs and overlap functions.
We have demonstrated the effect of continuum coupling in SFs and the overlap functions of drip line helium isotopes and isotones via comparing the results from GSM and NCSM calculations.
Added to that, the overlap functions obtained in GSM are all extended in space, whereas they are localized in NCSM. Isospin symmetry breaking is also clearly seen in GSM, as SFs and overlap functions of mirror nuclei vary significantly.
These features cannot be seen in NCSM overlap functions, even though the Coulomb force is included.
Finally, the spectroscopic factor and overlap function of $^6$He$(0_1^+)$ $\otimes \nu p_{3/2} \to $ $^7$He$(3/2_1^-)$ from GSM are compared with \textit{ab initio} no-core shell model, variational Monte Carlo, and Green’s function Monte
Carlo calculations. The results give that the asymptotic behavior of the nuclear wave function, particularly for the resonance states in the drip line region, is well described in GSM calculations.

This study has then proved that continuum coupling and a proper depiction of many-body nuclear wave function asymptotes is prominent for the comprehension of drip line nuclei.
This provides the possibility to precisely study ANCs in drip line nuclei in the future, which are important for the calculation of capture cross sections at astrophysical energies, for example.

\textit{Acknowledgments.}~
 This work has been supported by the National Natural Science Foundation of China under Grant Nos.  12205340, 12175281, and 11975282;  the Gansu Natural Science Foundation under Grant No. 22JR5RA123;  the Strategic Priority Research Program of Chinese Academy of Sciences under Grant No. XDB34000000; the Key Research Program of the Chinese Academy of Sciences under Grant No. XDPB15; the State Key Laboratory of Nuclear Physics and Technology, Peking University under Grant No. NPT2020KFY13. This research was made possible by using the computing resources of Gansu Advanced Computing Center. 



\section*{References}

\bibliographystyle{elsarticle-num_noURL}
\bibliography{Ref}

\begin{thebibliography}{10}
\expandafter\ifx\csname url\endcsname\relax
  \def\url#1{\texttt{#1}}\fi
\expandafter\ifx\csname urlprefix\endcsname\relax\def\urlprefix{URL }\fi
\expandafter\ifx\csname href\endcsname\relax
  \def\href#1#2{#2} \def\path#1{#1}\fi

\bibitem{SMIRNOVA2010109}
N.~Smirnova, B.~Bally, K.~Heyde, F.~Nowacki, K.~Sieja, \href{https://www.sciencedirect.com/science/article/pii/S0370269310002388}{Phys. Lett. B} 686~(2) (2010) 109--113.

\bibitem{PhysRevLett.109.032502}
G.~Hagen, M.~Hjorth-Jensen, G.~R. Jansen, R.~Machleidt, T.~Papenbrock, \href{https://link.aps.org/doi/10.1103/PhysRevLett.109.032502}{Phys. Rev. Lett.} 109 (2012) 032502.

\bibitem{RevModPhys.76.215}
A.~S. Jensen, K.~Riisager, D.~V. Fedorov, E.~Garrido, \href{https://link.aps.org/doi/10.1103/RevModPhys.76.215}{Rev. Mod. Phys.} 76 (2004) 215--261.

\bibitem{PhysRevLett.116.212501}
J.~Bonnard, S.~M. Lenzi, A.~P. Zuker, \href{https://link.aps.org/doi/10.1103/PhysRevLett.116.212501}{Phys. Rev. Lett.} 116 (2016) 212501.

\bibitem{RevModPhys.84.567}
M.~Pf\"utzner, M.~Karny, L.~V. Grigorenko, K.~Riisager, \href{https://link.aps.org/doi/10.1103/RevModPhys.84.567}{Rev. Mod. Phys.} 84 (2012) 567--619.

\bibitem{DETRAZ1980307}
C.~Detraz, M.~Epherre, D.~Guillemaud, P.~Hansen, B.~Jonson, R.~Klapisch, M.~Langevin, S.~Mattsson, F.~Naulin, G.~Nyman, A.~Poskanzer, H.~Ravn, M.~{de Saint-Simon}, K.~Takahashi, C.~Thibault, F.~Touchard, \href{https://www.sciencedirect.com/science/article/pii/0370269380908849}{Phys. Lett. B} 94~(3) (1980) 307--309.

\bibitem{KOURA2015228}
H.~Koura, S.~Chiba, \href{https://www.sciencedirect.com/science/article/pii/S1876610214027039}{Energy Procedia} 71 (2015) 228--236, the Fourth International Symposium on Innovative Nuclear Energy Systems, INES-4.

\bibitem{PhysRevLett.43.1652}
R.~E. Azuma, L.~C. Carraz, P.~G. Hansen, B.~Jonson, K.~L. Kratz, S.~Mattsson, G.~Nyman, H.~Ohm, H.~L. Ravn, A.~Schr\"oder, W.~Ziegert, \href{https://link.aps.org/doi/10.1103/PhysRevLett.43.1652}{Phys. Rev. Lett.} 43 (1979) 1652--1654.

\bibitem{PhysRevC.64.061301}
F.~M. Marqu\'es, M.~Labiche, N.~A. Orr, J.~C. Ang\'elique, L.~Axelsson, B.~Benoit, U.~C. Bergmann, M.~J.~G. Borge, W.~N. Catford, S.~P.~G. Chappell, N.~M. Clarke, G.~Costa, N.~Curtis, A.~D'Arrigo, E.~d.~G. Brennand, F.~d.~O. Santos, O.~Dorvaux, G.~Fazio, M.~Freer, B.~R. Fulton, G.~Giardina, S.~Gr\'evy, D.~Guillemaud-Mueller, F.~Hanappe, B.~Heusch, B.~Jonson, C.~L. Brun, S.~Leenhardt, M.~Lewitowicz, M.~J. L\'opez, K.~Markenroth, A.~C. Mueller, T.~Nilsson, A.~Ninane, G.~Nyman, I.~Piqueras, K.~Riisager, M.~G. Saint~Laurent, F.~Sarazin, S.~M. Singer, O.~Sorlin, L.~Stuttg\'e, \href{https://link.aps.org/doi/10.1103/PhysRevC.64.061301}{Phys. Rev. C} 64 (2001) 061301.

\bibitem{BAYE2011464}
D.~Baye, E.~Tursunov, \href{https://www.sciencedirect.com/science/article/pii/S0370269311000281}{Phys. Lett. B} 696~(5) (2011) 464--467.

\bibitem{PhysRevLett.124.212503}
K.~J. Cook, T.~Nakamura, Y.~Kondo, K.~Hagino, K.~Ogata, A.~T. Saito, N.~L. Achouri, T.~Aumann, H.~Baba, F.~Delaunay, Q.~Deshayes, P.~Doornenbal, N.~Fukuda, J.~Gibelin, J.~W. Hwang, N.~Inabe, T.~Isobe, D.~Kameda, D.~Kanno, S.~Kim, N.~Kobayashi, T.~Kobayashi, T.~Kubo, S.~Leblond, J.~Lee, F.~M. Marqu\'es, R.~Minakata, T.~Motobayashi, K.~Muto, T.~Murakami, D.~Murai, T.~Nakashima, N.~Nakatsuka, A.~Navin, S.~Nishi, S.~Ogoshi, N.~A. Orr, H.~Otsu, H.~Sato, Y.~Satou, Y.~Shimizu, H.~Suzuki, K.~Takahashi, H.~Takeda, S.~Takeuchi, R.~Tanaka, Y.~Togano, J.~Tsubota, A.~G. Tuff, M.~Vandebrouck, K.~Yoneda, \href{https://link.aps.org/doi/10.1103/PhysRevLett.124.212503}{Phys. Rev. Lett.} 124 (2020) 212503.

\bibitem{PhysRevLett.112.242501}
N.~Kobayashi, T.~Nakamura, Y.~Kondo, J.~A. Tostevin, Y.~Utsuno, N.~Aoi, H.~Baba, R.~Barthelemy, M.~A. Famiano, N.~Fukuda, N.~Inabe, M.~Ishihara, R.~Kanungo, S.~Kim, T.~Kubo, G.~S. Lee, H.~S. Lee, M.~Matsushita, T.~Motobayashi, T.~Ohnishi, N.~A. Orr, H.~Otsu, T.~Otsuka, T.~Sako, H.~Sakurai, Y.~Satou, T.~Sumikama, H.~Takeda, S.~Takeuchi, R.~Tanaka, Y.~Togano, K.~Yoneda, \href{https://link.aps.org/doi/10.1103/PhysRevLett.112.242501}{Phys. Rev. Lett.} 112 (2014) 242501.

\bibitem{PhysRevLett.89.042501}
R.~Id~Betan, R.~J. Liotta, N.~Sandulescu, T.~Vertse, \href{https://link.aps.org/doi/10.1103/PhysRevLett.89.042501}{Phys. Rev. Lett.} 89 (2002) 042501.

\bibitem{PhysRevC.105.L051301}
V.~Girard-Alcindor, A.~Mercenne, I.~Stefan, F.~de~Oliveira~Santos, N.~Michel, M.~P\l{}oszajczak, M.~Assi\'e, A.~Lemasson, E.~Cl\'ement, F.~Flavigny, A.~Matta, D.~Ramos, M.~Rejmund, J.~Dudouet, D.~Ackermann, P.~Adsley, M.~Assun\ifmmode \mbox{\c{c}}\else~\c{c}\fi{}\ ao, B.~Bastin, D.~Beaumel, G.~Benzoni, R.~Borcea, A.~J. Boston, D.~Brugnara, L.~C\'aceres, B.~Cederwall, I.~Celikovic, V.~Chudoba, M.~Ciemala, J.~Collado, F.~C.~L. Crespi, G.~D'Agata, G.~De~France, F.~Delaunay, C.~Diget, C.~Domingo-Pardo, J.~Eberth, C.~Foug\`eres, S.~Franchoo, F.~Galtarossa, A.~Georgiadou, J.~Gibelin, S.~Giraud, V.~Gonz\'alez, N.~Goyal, A.~Gottardo, J.~Goupil, S.~Gr\'evy, V.~Guimaraes, F.~Hammache, L.~J. Harkness-Brennan, H.~Hess, N.~Jovan\ifmmode \check{c}\else \v{c}\fi{}evi\ifmmode~\acute{c}\else \'{c}\fi{}, D.~S. Judson~Oliver, O.~Kamalou, A.~Kamenyero, J.~Kiener, W.~Korten, S.~Koyama, M.~Labiche, L.~Lalanne, V.~Lapoux, S.~Leblond, A.~Lefevre, C.~Lenain, S.~Leoni, H.~Li, A.~Lopez-Martens, A.~Maj, I.~Matea, R.~Menegazzo,
  D.~Mengoni, A.~Meyer, B.~Million, B.~Monteagudo, P.~Morfouace, J.~Mrazek, M.~Niikura, J.~Piot, Z.~Podolyak, C.~Portail, A.~Pullia, B.~Quintana, F.~Recchia, P.~Reiter, K.~Rezynkina, T.~Roger, J.~S. Rojo, F.~Rotaru, M.~D. Salsac, A.~M. S\'anchez~Ben\'{\i}tez, E.~Sanchis, M.~\ifmmode~\mbox{\c{S}}\else \c{S}\fi{}enyigit, N.~de~S\'er\'eville, M.~Siciliano, J.~Simpson, D.~Sohler, O.~Sorlin, M.~Stanoiu, C.~Stodel, D.~Suzuki, C.~Theisen, D.~Thisse, J.~C.Thomas, P.~Ujic, J.~J. Valiente-Dob\'on, M.~Zieli\ifmmode~\acute{n}\else \'{n}\fi{}ska, \href{https://link.aps.org/doi/10.1103/PhysRevC.105.L051301}{Phys. Rev. C} 105 (2022) L051301.

\bibitem{PhysRevC.72.054322}
R.~Id~Betan, R.~J. Liotta, N.~Sandulescu, T.~Vertse, R.~Wyss, \href{https://link.aps.org/doi/10.1103/PhysRevC.72.054322}{Phys. Rev. C} 72 (2005) 054322.

\bibitem{PhysRevC.67.054311}
N.~Michel, W.~Nazarewicz, M.~P\l{}oszajczak, J.~Oko\l{}owicz, \href{https://link.aps.org/doi/10.1103/PhysRevC.67.054311}{Phys. Rev. C} 67 (2003) 054311.

\bibitem{PhysRevC.100.064303}
N.~Michel, J.~G. Li, F.~R. Xu, W.~Zuo, \href{https://link.aps.org/doi/10.1103/PhysRevC.100.064303}{Phys. Rev. C} 100 (2019) 064303.

\bibitem{ZHANG2022136958}
S.~Zhang, Y.~Z. Ma, J.~G. Li, B.~S. Hu, Q.~Yuan, Z.~H. Cheng, F.~R. Xu, \href{https://www.sciencedirect.com/science/article/pii/S0370269322000922}{Phys. Lett. B} 827 (2022) 136958.

\bibitem{PhysRevC.104.024319}
J.~G. Li, N.~Michel, W.~Zuo, F.~R. Xu, \href{https://link.aps.org/doi/10.1103/PhysRevC.104.024319}{Phys. Rev. C} 104 (2021) 024319.

\bibitem{PhysRevC.103.034305}
J.~G. Li, N.~Michel, W.~Zuo, F.~R. Xu, \href{https://link.aps.org/doi/10.1103/PhysRevC.103.034305}{Phys. Rev. C} 103 (2021) 034305.

\bibitem{PhysRevLett.127.262502}
Y.~Jin, C.~Y. Niu, K.~W. Brown, Z.~H. Li, H.~Hua, A.~K. Anthony, J.~Barney, R.~J. Charity, J.~Crosby, D.~Dell'Aquila, J.~M. Elson, J.~Estee, M.~Ghazali, G.~Jhang, J.~G. Li, W.~G. Lynch, N.~Michel, L.~G. Sobotka, S.~Sweany, F.~C.~E. Teh, A.~Thomas, C.~Y. Tsang, M.~B. Tsang, S.~M. Wang, H.~Y. Wu, C.~X. Yuan, K.~Zhu, \href{https://link.aps.org/doi/10.1103/PhysRevLett.127.262502}{Phys. Rev. Lett.} 127 (2021) 262502.

\bibitem{PhysRevC.85.064320}
J.~Oko\l{}owicz, N.~Michel, W.~Nazarewicz, M.~P\l{}oszajczak, \href{https://link.aps.org/doi/10.1103/PhysRevC.85.064320}{Phys. Rev. C} 85 (2012) 064320.

\bibitem{PhysRevC.75.031301}
N.~Michel, W.~Nazarewicz, M.~P\l{}oszajczak, \href{https://link.aps.org/doi/10.1103/PhysRevC.75.031301}{Phys. Rev. C} 75 (2007) 031301.

\bibitem{PhysRevC.82.044315}
N.~Michel, W.~Nazarewicz, M.~P\l{}oszajczak, \href{https://link.aps.org/doi/10.1103/PhysRevC.82.044315}{Phys. Rev. C} 82 (2010) 044315.

\bibitem{XIE2023137800}
M.~R. Xie, J.~G. Li, N.~Michel, H.~H. Li, S.~T. Wang, H.~J. Ong, W.~Zuo, \href{https://www.sciencedirect.com/science/article/pii/S037026932300134X}{Phys. Lett. B} (2023) 137800.

\bibitem{PhysRevC.104.L061301}
J.~Wylie, J.~Oko\l{}owicz, W.~Nazarewicz, M.~P\l{}oszajczak, S.~M. Wang, X.~Mao, N.~Michel, \href{https://link.aps.org/doi/10.1103/PhysRevC.104.L061301}{Phys. Rev. C} 104 (2021) L061301.

\bibitem{OKOLOWICZ2003271}
J.~Okołowicz, M.~Płoszajczak, I.~Rotter, \href{https://www.sciencedirect.com/science/article/pii/S0370157302003666}{Phys. Rep.} 374~(4) (2003) 271--383.

\bibitem{Blank_2008}
B.~Blank, M.~Płoszajczak, \href{https://dx.doi.org/10.1088/0034-4885/71/4/046301}{Rep. Prog. Phys.} 71~(4) (2008) 046301.

\bibitem{PhysRevLett.95.042503}
J.~Rotureau, J.~Oko\l{}owicz, M.~P\l{}oszajczak, \href{https://link.aps.org/doi/10.1103/PhysRevLett.95.042503}{Phys. Rev. Lett.} 95 (2005) 042503.

\bibitem{BERGGREN1968265}
T.~Berggren, \href{http://www.sciencedirect.com/science/article/pii/0375947468905939}{Nucl. Phys. A} 109~(2) (1968) 265 -- 287.

\bibitem{0954-3899-36-1-013101}
N.~Michel, W.~Nazarewicz, M.~P{\l}oszajczak, T.~Vertse, \href{http://stacks.iop.org/0954-3899/36/i=1/a=013101}{J. Phys. G. Nucl. Part. Phys.} 36~(1) (2009) 013101.

\bibitem{physics3040062}
J.~G. Li, Y.~Z. Ma, N.~Michel, B.~S. Hu, Z.~H. Sun, W.~Zuo, F.~R. Xu, \href{https://www.mdpi.com/2624-8174/3/4/62}{Physics} 3~(4) (2021) 977--997.

\bibitem{Michel_Springer}
N.~Michel, M.~P{\l}oszajczak, Gamow Shell Model, The Unified Theory of Nuclear Structure and Reactions, Lecture Notes in Physics, Vol. 983, Springer, Berlin, 2021.

\bibitem{PhysRevC.100.054313}
J.~G. Li, N.~Michel, B.~S. Hu, W.~Zuo, F.~R. Xu, \href{https://link.aps.org/doi/10.1103/PhysRevC.100.054313}{Phys. Rev. C} 100 (2019) 054313.

\bibitem{PhysRevC.103.044319}
N.~Michel, J.~G. Li, F.~R. Xu, W.~Zuo, \href{https://link.aps.org/doi/10.1103/PhysRevC.103.044319}{Phys. Rev. C} 103 (2021) 044319.

\bibitem{PhysRevC.104.L061306}
H.~H. Li, J.~G. Li, N.~Michel, W.~Zuo, \href{https://link.aps.org/doi/10.1103/PhysRevC.104.L061306}{Phys. Rev. C} 104 (2021) L061306.

\bibitem{PhysRevLett.95.222501}
M.~B. Tsang, J.~Lee, W.~G. Lynch, \href{https://link.aps.org/doi/10.1103/PhysRevLett.95.222501}{Phys. Rev. Lett.} 95 (2005) 222501.

\bibitem{doi:10.1146/annurev.nucl.53.041002.110406}
P.~Hansen, J.~Tostevin, \href{https://doi.org/10.1146/annurev.nucl.53.041002.110406}{Ann. Rev. Nucl. Part. Sci.} 53~(1) (2003) 219--261.

\bibitem{PhysRevC.105.024613}
J.~Li, C.~A. Bertulani, F.~Xu, \href{https://link.aps.org/doi/10.1103/PhysRevC.105.024613}{Phys. Rev. C} 105 (2022) 024613.

\bibitem{PhysRevC.90.057602}
J.~A. Tostevin, A.~Gade, \href{https://link.aps.org/doi/10.1103/PhysRevC.90.057602}{Phys. Rev. C} 90 (2014) 057602.

\bibitem{PhysRevC.103.054610}
J.~A. Tostevin, A.~Gade, \href{https://link.aps.org/doi/10.1103/PhysRevC.103.054610}{Phys. Rev. C} 103 (2021) 054610.

\bibitem{PhysRevC.70.064313}
N.~Michel, W.~Nazarewicz, M.~P\l{}oszajczak, \href{https://link.aps.org/doi/10.1103/PhysRevC.70.064313}{Phys. Rev. C} 70 (2004) 064313.

\bibitem{BARRETT2013131}
B.~R. Barrett, P.~Navrátil, J.~P. Vary, \href{http://www.sciencedirect.com/science/article/pii/S0146641012001184}{Prog. Part. Nucl. Phys.} 69 (2013) 131 -- 181.

\bibitem{PhysRev.56.750}
A.~J.~F. Siegert, \href{https://link.aps.org/doi/10.1103/PhysRev.56.750}{Phys. Rev.} 56 (1939) 750--752.

\bibitem{PhysRevLett.89.042502}
N.~Michel, W.~Nazarewicz, M.~P\l{}oszajczak, K.~Bennaceur, \href{https://link.aps.org/doi/10.1103/PhysRevLett.89.042502}{Phys. Rev. Lett.} 89 (2002) 042502.

\bibitem{ensdf}
\url{http://www.nndc.bnl.gov/ensdf}.

\bibitem{TILLEY20023}
D.~Tilley, C.~Cheves, J.~Godwin, G.~Hale, H.~Hofmann, J.~Kelley, C.~Sheu, H.~Weller, \href{https://www.sciencedirect.com/science/article/pii/S0375947402005973}{Nucl. Phys. A} 708~(1) (2002) 3--163.

\bibitem{PhysRevC.84.014320}
R.~J. Charity, J.~M. Elson, J.~Manfredi, R.~Shane, L.~G. Sobotka, B.~A. Brown, Z.~Chajecki, D.~Coupland, H.~Iwasaki, M.~Kilburn, J.~Lee, W.~G. Lynch, A.~Sanetullaev, M.~B. Tsang, J.~Winkelbauer, M.~Youngs, S.~T. Marley, D.~V. Shetty, A.~H. Wuosmaa, T.~K. Ghosh, M.~E. Howard, \href{https://link.aps.org/doi/10.1103/PhysRevC.84.014320}{Phys. Rev. C} 84 (2011) 014320.

\bibitem{PhysRevC.82.041304}
R.~J. Charity, J.~M. Elson, J.~Manfredi, R.~Shane, L.~G. Sobotka, Z.~Chajecki, D.~Coupland, H.~Iwasaki, M.~Kilburn, J.~Lee, W.~G. Lynch, A.~Sanetullaev, M.~B. Tsang, J.~Winkelbauer, M.~Youngs, S.~T. Marley, D.~V. Shetty, A.~H. Wuosmaa, T.~K. Ghosh, M.~E. Howard, \href{https://link.aps.org/doi/10.1103/PhysRevC.82.041304}{Phys. Rev. C} 82 (2010) 041304.

\bibitem{10.1143/PTP.62.981}
H.~Furutani, H.~Horiuchi, R.~Tamagaki, \href{https://doi.org/10.1143/PTP.62.981}{Prog. Theo. Phys.} 62~(4) (1979) 981--1002.

\bibitem{PhysRevC.96.054316}
Y.~Jaganathen, R.~M.~I. Betan, N.~Michel, W.~Nazarewicz, M.~P\l{}oszajczak, \href{https://link.aps.org/doi/10.1103/PhysRevC.96.054316}{Phys. Rev. C} 96 (2017) 054316.

\bibitem{PhysRevC.106.L011301}
N.~Michel, J.~G. Li, L.~H. Ru, W.~Zuo, \href{https://link.aps.org/doi/10.1103/PhysRevC.106.L011301}{Phys. Rev. C} 106 (2022) L011301.

\bibitem{PhysRevC.91.034609}
K.~Fossez, N.~Michel, M.~P\l{}oszajczak, Y.~Jaganathen, R.~M. Id~Betan, \href{https://link.aps.org/doi/10.1103/PhysRevC.91.034609}{Phys. Rev. C} 91 (2015) 034609.

\bibitem{SHIROKOV201687}
A.~Shirokov, I.~Shin, Y.~Kim, M.~Sosonkina, P.~Maris, J.~Vary, \href{https://www.sciencedirect.com/science/article/pii/S0370269316304269}{Phys. Lett. B} 761 (2016) 87--91.

\bibitem{PhysRevC.99.054308}
G.~A. Negoita, J.~P. Vary, G.~R. Luecke, P.~Maris, A.~M. Shirokov, I.~J. Shin, Y.~Kim, E.~G. Ng, C.~Yang, M.~Lockner, G.~M. Prabhu, \href{https://link.aps.org/doi/10.1103/PhysRevC.99.054308}{Phys. Rev. C} 99 (2019) 054308.

\bibitem{Shin_2017}
I.~J. Shin, Y.~Kim, P.~Maris, J.~P. Vary, C.~Forss{\'{e}}n, J.~Rotureau, N.~Michel, \href{https://doi.org/10.1088/1361-6471/aa6cb7}{Jour. Phys. G: Nucl. Part. Phys.} 44~(7) (2017) 075103.

\bibitem{LI2022137225}
J.~G. Li, N.~Michel, H.~H. Li, W.~Zuo, \href{https://www.sciencedirect.com/science/article/pii/S0370269322003598}{Phys. Lett. B} 832 (2022) 137225.

\bibitem{PhysRevC.84.024319}
I.~Brida, S.~C. Pieper, R.~B. Wiringa, \href{https://link.aps.org/doi/10.1103/PhysRevC.84.024319}{Phys. Rev. C} 84 (2011) 024319.

\bibitem{CAO201246}
Z.~X. Cao, Y.~L. Ye, J.~Xiao, L.~H. Lv, D.~X. Jiang, T.~Zheng, H.~Hua, Z.~H. Li, X.~Q. Li, Y.~C. Ge, J.~L. Lou, R.~Qiao, Q.~T. Li, H.~B. You, R.~J. Chen, D.~Y. Pang, H.~Sakurai, H.~Otsu, M.~Nishimura, S.~Sakaguchi, H.~Baba, Y.~Togano, K.~Yoneda, C.~Li, S.~Wang, H.~Wang, K.~A. Li, T.~Nakamura, Y.~Nakayama, Y.~Kondo, S.~Deguchi, Y.~Satou, K.~Tshoo, \href{https://www.sciencedirect.com/science/article/pii/S0370269311014523}{Phys. Lett. B} 707~(1) (2012) 46--51.

\bibitem{BECK2009496}
F.~Beck, D.~Frekers, P.~{von Neumann-Cosel}, A.~Richter, N.~Ryezayeva, I.~Thompson, \href{https://www.sciencedirect.com/science/article/pii/S0370269308015402}{Phys. Lett. B} 671~(4) (2009) 496.

\bibitem{PhysRevC.78.041302}
A.~H. Wuosmaa, J.~P. Schiffer, K.~E. Rehm, J.~P. Greene, D.~J. Henderson, R.~V.~F. Janssens, C.~L. Jiang, L.~Jisonna, J.~C. Lighthall, S.~T. Marley, E.~F. Moore, R.~C. Pardo, N.~Patel, M.~Paul, D.~Peterson, S.~C. Pieper, G.~Savard, R.~E. Segel, R.~H. Siemssen, X.~D. Tang, R.~B. Wiringa, \href{https://link.aps.org/doi/10.1103/PhysRevC.78.041302}{Phys. Rev. C} 78 (2008) 041302.

\bibitem{PhysRevLett.125.192503}
J.~Lee, X.~X. Xu, K.~Kaneko, Y.~Sun, C.~J. Lin, L.~J. Sun, P.~F. Liang, Z.~H. Li, J.~Li, H.~Y. Wu, D.~Q. Fang, J.~S. Wang, Y.~Y. Yang, C.~X. Yuan, Y.~H. Lam, Y.~T. Wang, K.~Wang, J.~G. Wang, J.~B. Ma, J.~J. Liu, P.~J. Li, Q.~Q. Zhao, L.~Yang, N.~R. Ma, D.~X. Wang, F.~P. Zhong, S.~H. Zhong, F.~Yang, H.~M. Jia, P.~W. Wen, M.~Pan, H.~L. Zang, X.~Wang, C.~G. Wu, D.~W. Luo, H.~W. Wang, C.~Li, C.~Z. Shi, M.~W. Nie, X.~F. Li, H.~Li, P.~Ma, Q.~Hu, G.~Z. Shi, S.~L. Jin, M.~R. Huang, Z.~Bai, Y.~J. Zhou, W.~H. Ma, F.~F. Duan, S.~Y. Jin, Q.~R. Gao, X.~H. Zhou, Z.~G. Hu, M.~Wang, M.~L. Liu, R.~F. Chen, X.~W. Ma, \href{https://link.aps.org/doi/10.1103/PhysRevLett.125.192503}{Phys. Rev. Lett.} 125 (2020) 192503.

\bibitem{PhysRevLett.109.102501}
Y.~H. Zhang, H.~S. Xu, Y.~A. Litvinov, X.~L. Tu, X.~L. Yan, S.~Typel, K.~Blaum, M.~Wang, X.~H. Zhou, Y.~Sun, B.~A. Brown, Y.~J. Yuan, J.~W. Xia, J.~C. Yang, G.~Audi, X.~C. Chen, G.~B. Jia, Z.~G. Hu, X.~W. Ma, R.~S. Mao, B.~Mei, P.~Shuai, Z.~Y. Sun, S.~T. Wang, G.~Q. Xiao, X.~Xu, T.~Yamaguchi, Y.~Yamaguchi, Y.~D. Zang, H.~W. Zhao, T.~C. Zhao, W.~Zhang, W.~L. Zhan, \href{https://link.aps.org/doi/10.1103/PhysRevLett.109.102501}{Phys. Rev. Lett.} 109 (2012) 102501.

\end{thebibliography}





\end{document}